\documentclass[eprint, twocolumn]{revtex4-2}
\usepackage{bbm}
\usepackage{natbib}
\usepackage{graphics}
\usepackage{amsmath}
\usepackage{mathrsfs}
\usepackage{theorem}
\usepackage{float}
\usepackage{hyperref}
\usepackage{rotating}
\usepackage{amssymb}
\usepackage[english]{babel}
\usepackage{color}
\usepackage{fancybox}

\newcommand{\ket}[1]{|#1\rangle}

\newcommand{\be}{\begin{eqnarray}}
\newcommand{\ee}{\end{eqnarray}}

\begin{document}

\title{Quantum Mpemba effect of localization in the dissipative  mosaic model   }
\author{J. W. Dong}
\author{H. F. Mu}
\author{M. Qin}
\author{H. T. Cui}
\email{cuiht01335@aliyun.com}
\affiliation{School of Physics and Optoelectronic Engineering and Institute of Theoretical Physics, Ludong University, Yantai 264025, China}
\date{\today}

\begin{abstract}
The quantum Mpemba effect in open quantum systems has been extensively studied, but a comprehensive understanding of this phenomenon remains elusive. In this paper, we conduct an analytical investigation of the dissipative dynamics of single excitations in the The mosaic model. Surprisingly,  we discover that the presence of an asymptotic mobility edge, denoted as $E_c^{\infty}$, can lead to  unique dissipation behavior,  serving as a hallmark of the quantum Mpemba effect. Especially, it is found that the energy level $E_c^{\infty}$ exhibits a global periodicity in the real configuration, which acts to inhibit dissipation in the system. Conversely, when the system deviates from $E_c^{\infty}$, the quasidisorder sets in, leading to increased dissipative effects due to the broken of periodicity. Furthermore, we find that the rate of dissipation is closely linked to the localization of the initial state. As a result,  the quantum  Mpemba effect can be observed clearly by a measure of localization.
\end{abstract}

\maketitle

\section{introduction}
The Mpemba effect (ME) has had a long history of research since its first experimental discovery in 1963\cite{mpemba}. In classical systems, the ME shows that a thermal system at a high temperature $T_h$ can be cooled down  to equilibrium  faster than one at a low temperature $T_l (< T_h)$ under identical condition. In the quantum realm, the quantum Mpemba effect (QME) characterizes the feature that a quantum system far from equilibrium can relax to equilibrium faster than a system  closer to equilibrium. Different methods are used to study the QME, depending on the properties of the system of interest. In closed integrable systems, entanglement asymmetry is used to measure the restoration of symmetry in initially symmetry-broken states \cite{ares23a, murciano24, ares23b, chalas24, rylands24, joshi24, mbl24, liu24}.  The QME can also be observed in  quantum simulations, in which systems starting with higher energy can relax  faster than those starting with low energy\cite{chang24}.

Another intriguing situation arises in  open quantum systems, in which  the dissipation  can exhibit significant dependence on the initial condition \cite{lu17, carollo, kochsiek, zhang24, strachan, nava, wang, akc23, akc24, wsw24}. For instance,  it was recently found that the approach to the stationary state  in  Markovian open quantum systems  can be exponentially accelerated  by performing a unitary transformation \cite{carollo, kochsiek}.  This phenomenon is known as the strong ME, which has been verified experimentally \cite{zhang24}. Furthermore, the QME can also be observed in open nonequilibrium quantum systems, where the system can  relax to nonequilibrium steady states (NESS). By  measuring  the distance to NESS, the QME can be identified by the intersection point of evolution paths at a finite time for distinct initial states  \cite{nava, wang}. Another interesting  the QME was introduced recently, that  involves  memory time scales induced by the  non-Markovian  dynamics of an open quantum system \cite{strachan}. Despite of the observation of the QME in various systems, a general understanding of it remains elusive.

In this paper, we introduce a QME related to the localization in the  The mosaic model coupled to a bosonic environment. The  mosaic model describes the one-dimensional quasiperiodic lattice system with the sinusoidal disorder appearing on the sites mediated by a unit cell composed of  $\kappa$ sublattices. The mobility edge can be emergent due to the competition between the  disorder and periodicity of unit cell. Crucially, the asymptotic mobility edge can be identified, independent of the strength of disorder.  As shown in the following discussion, the dissipation of a single excitation can demonstrate the significant correlation to the localization in an initial state  due to the presence of asymptotic mobility edge. In contrast to  the previous studies, the dissipative dynamics of excitation can be evaluated analytically in this case using  Laplace transformation.  The exact determination of the effective modes dominating the dynamics provides clear evidence of the QME of localization.



\section{Model and method }
The  mosaic model coupled to a bosonic environment is depicted  by the  total Hamiltonian
\be
H= H_{s} + H_b + H_{int}.
\ee
$H_s$ depicts the  mosaic model \cite{the mosaic}, written as
\be\label{hs}
& H_s=\sum_{n=1}^N \lambda  \left(c^{\dagger}_{n} c_{n+1} +c^{\dagger}_{n+1} c_{n} \right) +
\Delta_n c^{\dagger}_n c_{n}, \\
& \Delta_n=\begin{cases}
                    \Delta  \cos\left (2\pi \beta n +\phi\right), & \mbox{if } n=j\kappa (j=1, 2, \cdots) \\
                    0, & \mbox{otherwise}.
                  \end{cases} \nonumber
\ee
where $N$ denotes the number of lattice and $c_n (c^{\dagger}_n)$ is the annihilation (creation) operator of the excitation at the $n$th  site. $\lambda$ characterizes the hopping strength, which is assumed to be unitary in the following.  $\beta=\left(\sqrt{5}-1\right)/2$ is the golden ratio, which is responsible for the quasiperiodicity in the system. $\kappa$ is a positive integer, which denotes the size of the unit cell.  When $\kappa=1$, the   mosaic model is reduced to the   Abury-Andr\'e-Harper model \cite{aa,haper}, for which  the duality symmetry leads to the occurrence of  localization-delocalization transition \cite{aa,haper}.  However, for $\kappa \neq 1$, this symmetry is broken. Thus, the mobility edge (ME) $E_c$  can be determined  exactly by \cite{the mosaic}
\be\label{me}
&\left|\frac{\Delta}{2} a_k \right| = 1,  \nonumber \\
& a_k= \frac{1}{\sqrt{E_c^2-4}} \left[ \left(\frac{E_c +\sqrt{E_c^2-4} }{2}\right)^{\kappa}- \left(\frac{E_c -\sqrt{E_c^2-4} }{2}\right)^{\kappa} \right]
\ee
which separates the extended energy levels from localized ones. In \ref{appendix:E}, the localization of the mosaic model is depicted in Fig. \ref{fig:E}  by the inverse  participation ratio. It is observed that as $\Delta\rightarrow \infty$,  $E_c$ converges to an asymptotic  value $E^{\infty}_c$, determined by solving $a_k=0$. For instance, when $\kappa=2$, we find that $E_c= \pm \frac{2}{\Delta}$ and $E^{\infty}_c=0$, while for $\kappa=3$, we have $E_c= \pm \sqrt{1\pm 2/\Delta}$ and $E^{\infty}_c= \pm 1$.  We will show in following that the energy level corresponding to $E^{\infty}_c$ can exhibit significant resilience to dissipation. Furthermore,  the stability of  $E^{\infty}_c$ arises from the inherent discrete translational invariance of the mosaic model. To avoid the boundary effect, the periodic boundary condition $c^{(\dagger)}_{N+n}=c^{(\dagger)}_n$ is imposed in the discussion.

$H_b$ depicts the environment,  written as
\be
H_b= \sum_k \omega_k b_k^{\dagger}b_k,
\ee
where $b_k$ ($b_k^{\dagger}$) is the bosonic annihilation (creation) operator for mode $k$. The coupling between the mosaic model and its environment is depicted by
\be\label{hint}
H_{int}= \sum_{k, n}\left(g_k b_k c_n^{+} + g_k^* b_k^{\dagger} c_n\right)
\ee
where $g_k$ are the coupling constants. The open dynamics of the mosaic model is determined  by the spectral density
\be
J(\omega)= \sum_k \left| g_k \right|^2 \delta\left(\omega- \omega_k\right).
\ee
For specification, we choose the Lorentzian spectral density \cite{bellomo} in the following calculation, which has the form
\be\label{lorentz}
J(\omega)= \frac{\eta \omega_c^2}{\omega^2 + \omega_c^2},
\ee
where $\omega_c$ is the spectrum width and $\eta$ is the coupling strength. $\omega$ is
confined  to the interval $\left(-\infty, +\infty \right)$, allowing for an unbounded spectrum in the environment.  This choice rules out the occurrence of discrete bound state  \cite{bs1, bs2}, which keeps excitation from dissipating by opening an energy gap between the bound state and the spectrum in environment. In this circumstances, the excitation may ultimately be absorbed by the environment,leading to a unique equilibrium in this scenario. 

Because of  the absence of particle interaction in the mosaic model, the study focuses on the dynamics for a single excitation at zero temperature. In this scenario,  the state of the  system and environment can be written in the compact form
\be\label{state}
\ket{\psi(t)}= \left(\sum_{n=1}^N \alpha_n(t) c^{\dagger}_n+ \sum_{k=1}^{K} \beta_k(t) b^{\dagger}_k \right)\ket{0},
\ee
where $\ket{0}$ is the vacuum state, and $K$ denotes the number of frequency modes in the environment. Substituting Eq. \eqref{state} into the Schr\"{o}dinger equation and solving first for $\beta_k (t)$, one can find an integrodifferential equation for $\alpha_n(t)$,
\be\label{evolution}
\mathbbm{i}\frac{\partial }{\partial t}\alpha_n(t)&=& \left[\alpha_{n+1}(t) + \alpha_{n-1}(t)\right]+ \Delta_n \alpha_n(t) \nonumber\\
&&- \mathbbm{i} \sum_{n=1}^N \int_0^t \text{d}\tau\alpha_n(\tau)f(t-\tau),
\ee
where $\mathbbm{i}= \sqrt{-1}$  and
\be\label{memory}
f(t-\tau)= \int_0^{\infty} \text{d} \omega J(\omega)e^{-\mathbbm{i} \omega(t-\tau)}.
\ee
is the memory kernel.

This equation can be solved both numerically and analytically.  The exact numerical evaluation involves discretizing the evolution time equally and finding $\alpha_n(t)$ through iteration. However, this method becomes very exhaustive for long evolution times.  Alternatively,  Laplace transformation is adopted to find the analytical expression of  $\alpha_n(t)$, showing its asymptotic properties. By Laplace transformation
\be
A_n(p)= \int_{0}^{\infty} \text{d} t e^{- p t} \alpha_n(t),
\ee
Eq. \eqref{evolution} can be rewritten as
\be\label{laplace}
\mathbbm{i} \left[p A_n(p)- \alpha_n(0) \right]&=& A_{n+1}(p) + A_n(p) + \Delta_n A_n(p)  \nonumber \\
&-& \mathbbm{i}\int_{- \infty}^{+\infty} \text{d}\omega \frac{J(\omega)}{p + \mathbbm{i}\omega}  \sum_m A_m(p),
\ee
which constitutes a linear system of equations for $A_n(p)$ and thus can be  solved using Cramer's rule. However, it is noted that the integral $\int_{- \infty}^{+\infty} \text{d}\omega \tfrac{J(\omega)}{p + \mathbbm{i}\omega}$ becomes divergent when $\mathbbm{i} p = \omega$. This problem can be resolved by virtue of the  Sokhotski-Plemdj (SP) formula
\be
\lim_{\epsilon\rightarrow 0} \frac{1}{x- x_0 - \mathbbm{i}\epsilon} = \text{P} \frac{1}{x -x_0}  + \mathbbm{i} \pi \delta \left( x- x_0\right),
\ee
where $\text{P}$ denotes the principle value.
One thus gets
\be
\lim_{\epsilon\rightarrow 0} \int_{-\infty}^{+\infty} \text{d}\omega \frac{J(\omega) }{ \omega- \mathbbm{i}p - \mathbbm{i}\epsilon} = \text{P} \int_{-\infty}^{+\infty} \text{d}\omega \frac{J(\omega) }{ \omega- \mathbbm{i}p} + \mathbbm{i}\pi J(\mathbbm{i}p).\nonumber
\ee
For instance,  one gets, like for the Lorentzian spectral density \eqref{lorentz}
\be
\int_{-\infty}^{+\infty} \text{d}\omega \frac{J(\omega) }{ \omega- z}= \frac{\eta \pi \omega_c}{\mathbbm{i}\omega_c + z}, \nonumber
\ee
where  $\mathbbm{i}p=z$ is assumed for convenience.

To determine $\alpha_n(t)$, the inverse Laplace transformation, defined as
\be\label{inverselaplace}
\alpha_n(t)=\frac{1}{2\pi \mathbbm{i}} \int_{s-\mathbbm{i}\infty}^{s+\mathbbm{i}\infty} \text{d}p A_n(p) e^{p t}
\ee
is applied. Obviously, $\mathbbm{i}p$  characterizes  the effective mode, which determines the single-excitation dynamics. Equation \eqref{inverselaplace} can be evaluated using the residue theorem. For this purpose, one has to find the poles of $A_n(p)$,  denoted as $\mathbbm{i}p_i=z_i$, which are  the zero points of the determinant of the coefficient matrix of Eq. \eqref{laplace}. Formally, one can get finally
\be\label{alphant}
\alpha_n(t)\approx \sum_i c_{n, i} e^{- \mathbbm{i}z_i t},
\ee
where $c_{n,i}$ corresponds to the residue of $A_n(p)$ at $z_i$. It is obvious that the set of poles $z_i$ consists of the dynamical modes, which determines the dissipation  of excitation. Especially, the imaginary part of $z_i$ gives rise to  the  decay rate of excitation, which is responsible for the QME observed in the recent works \cite{carollo, kochsiek, zhang24}.  We emphasize that the summation  $\sum_n \alpha_n^*(0) c_{n,i}$ cannot be considered to be  the  probability on the mode $z_i$  for the initial condition $\left\{\alpha_n(0)\right\}$ since the amount of $z_i$'s is always larger than the dimension of the Hilbert space in the system. Furthermore, it is found that $\sum_n \alpha_n^*(0) c_{n,i}$ can be complex, and moreover the modulus may be larger than unit.

\section{Dynamical mode and localization in the mosaic model }

\begin{table}[tb]
  \centering
  \begin{tabular}{c|c|c}
    \hline \hline
   $z_i$  & $\kappa=2$ & $\kappa=3$ \\
     \hline
    $z_1$ & 3.37164 - $\mathbbm{i}$ 0.193387 & 3.125206 - $\mathbbm{i}$ 0.243343 \\
    $z_2$ & 2.17811 - $\mathbbm{i}$ 0.0654878 & 1.896475 - $\mathbbm{i}$  0.036575\\
    $z_3$ & 1.77457 - $\mathbbm{i}$ 0.00509817 &  1.477545 - $\mathbbm{i}$ 0.002295\\
    $z_4$ & 1.22702 - $\mathbbm{i}$ 0.0119018  &\underline{ 1.0 -$\mathbbm{i}$  5.07 $\times 10^{-17}$}\\
    $z_5$ & 0.594323 - $\mathbbm{i}$ 0.00182872  & 0.774981 - $\mathbbm{i}$ 0.004798 \\
    $z_6$ & \underline{1.06$\times 10^{-16}$- $\mathbbm{i}$ 7.30$\times 10^{-17}$} & -0.126563 - $\mathbbm{i}$ 0.006496 \\
    $z_7$ & -0.0912537 - $\mathbbm{i} $0.00428128 & -0.371380 - $\mathbbm{i} $ 0.00004543 \\
    $z_8$ & -0.963238 - $\mathbbm{i} $0.00926696 & \underline{-1.0 - $\mathbbm{i}$ 7.16 $\times 10^{-17}$} \\
    $z_9$ & -1.13057 - $\mathbbm{i} $0.0016377 & -1.13493 - $\mathbbm{i} $ 0.67839  \\
    $z_{10}$ & -1.13266 - $\mathbbm{i} $0.608806 & -1.227166 - $\mathbbm{i} $ 0.0035706  \\
    $z_{11}$ & -1.69034 - $\mathbbm{i} $0.0030446 & -1.930205 - $\mathbbm{i} $ 0.00685497  \\
    $z_{12}$ & -2.61258 - $\mathbbm{i} $0.00225544 & -2.573718 - $\mathbbm{i} $ 0.0170292  \\
    $z_{13}$ & -2.84288 - $\mathbbm{i} $0.00960152 & -2.787129 - $\mathbbm{i} $ 0.000595899  \\
    \hline  \hline
  \end{tabular}
  \caption{The  poles $z$  solved by finding the zeros of the determinant of the coefficient matrix in Eq. \eqref{laplace}. The steady poles are highlighted by underline. For this evaluation, $N=12, \Delta=2, \phi=0, \eta=0.1$, and $\omega_c=1$ are chosen. }\label{z}
\end{table}

With these preparations, we are ready to investigate the single-excitation dynamics in the   mosaic model.
By setting the determinant of the coefficient matrix to  zero, one can derive an equation of degree $N+1$ for pole $z$. As an illustration,  the solved  poles are demonstrated  for the case of  $\kappa=2$ and $\kappa=3$ respectively  in Table \ref{z} with $\Delta=2$ and $N=12$. For $\kappa=2$, one has  $E_c=\pm 1$  and for $\kappa=3$,  $E_c=0, \pm\sqrt{2}$.  It is worth noting that all the poles have negative imaginary parts,  indicating the dissipative nature of the excitation in the system. An interesting observation is that the imaginary part of the pole is closely related to its corresponding real part. For instance, in the case of $\kappa=2$,  the pole $z_6$ has a vanishing imaginary part, suggesting that this dynamics mode is resistant to dissipation. Interestingly,  its real part tends towards zero, coinciding with the value of $E^{\infty}_c$ for $\kappa=2$. It should be emphasized that this coincidence  is not occasional, as a similar pattern can be seen for  poles $z_4$ and $z_8$ for $\kappa=3$, where their real parts match the value of $E^{\infty}_c$ for $\kappa=3$. Apart from the steady poles, the imaginary part of  the other poles show a significant increase, indicating varying speeds of dissipation depending on the distance from the steady pole. This suggests that excitations may exhibit different rates of dissipation based on their proximity to the steady pole. Consequently, the QME can be readily observed under these conditions.

\begin{figure}
\center
\includegraphics[width=6cm]{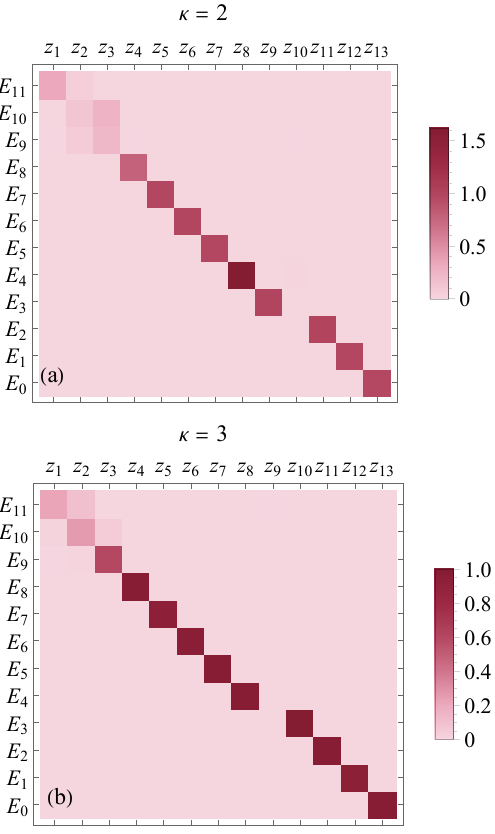}
\caption{(Color online) The square modulus for the overlap between the energy level $E_j (j=0, 1, 2, \cdots, 11)$ (labeled in descending order) in the   mosaic model  and the coefficient set $\left\{c_{n, i} \right\}(n=1, 2, \cdots, N)$ related to the initial state $\ket{E_j}$. For the plots, the parameters are chosen to be the same as those in Table \ref{z}.}
\label{fig:overlap}
\end{figure}

\begin{figure*}[tb]
\center
\includegraphics[width=15cm]{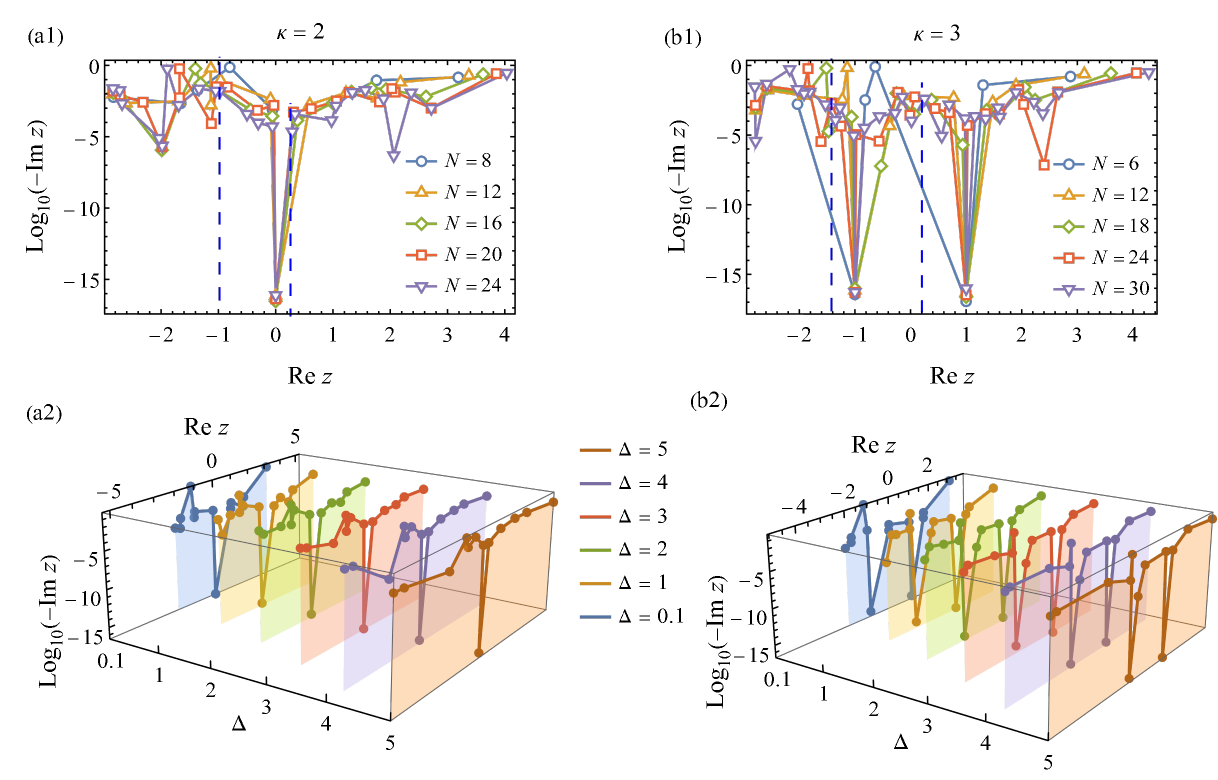}
\caption{(Color online) The plots for the poles $z_i$ with different $\Delta$'s and $N$'s. In panels (a1) and (b1), $\Delta=2$ is chosen. The blue-dashed lines labels the position of $\text{Re} z =E_c$. In panels (a2) and (b2),  $N=12$ is chosen. The other parameters are the same as those in Table \ref{z}. }
\label{fig:vN}
\end{figure*}

It is a natural assumption that the measure of "distance" is related to the localization in the mosaic model. To confirm this point,  the overlap between the energy level $E_j (j=0, 1, 2, \cdots, N-1)$  of the  mosaic model  and the corresponding  coefficient set $\left\{c_{n, i} \right\}(n=1, 2, \cdots, N)$ for  mode $z_i$, is shown in Fig. \ref{fig:overlap}. According to Eq. \eqref{alphant}, this overlap actually characterizes the contribution of mode $z_i$ to the survival probability of the energy eigenstate in the mosaic model for the initial state. As depicted in Fig. \ref{fig:overlap}, the energy level $E_j$ can  display  a significant overlap with the set  $\left\{c_{n, i} \right\}$ for a special  mode $z_i$. While  the overlap cannot  be interpreted as the probability of the initial state on the mode $z_i$, it does indicate the influence of the dynamic mode $z_i$ on the dissipative evolution of the energy level. This suggests that   a particular  mode $z_i$ will play a dominant role in the dissipation for a given level $E_j$ for the initial state. In this context,  mode $z_i$  may be seen as the renormalization of $E_j$. Since the level $E_j$ exhibits distinct localizations due to the presence of a mobility edge \cite{the mosaic}, the  dissipation characterized  by mode $z_i$ may be seen to be a  consequence of localization in the energy level. Therefore, one could  adopt the localization as a measure of the deviation of the initial state  from equilibrium. The dissipation leads to the system reaching equilibrium  by releasing excitation into the environment, ultimately causing the measure of localization to diminish..

Some comments should be made now. First as depicted in Fig. \ref{fig:vN},  the presence of steady poles $z_6$ or $z_{4(8)}$ is not influenced by the value of $N$ and $\Delta$, indicating that it is a inherent characteristic of the mosaic model.  To understand the physical basis for the steady poles, we have explicitly calculated the energy levels $E_6$ for $\kappa=2$ and $E_{4(8)}$ for $\kappa=3$  because of their close correspondence to the steady poles, as illustrated in Fig. \ref{fig:overlap}. It is found that the eigenstates  corresponding to  $E_6$ and $E_{4(8)}$  exhibit a distinct periodicity in the real configuration,  which can be expressed as
\be
\ket{E_6}&=& \frac{1}{\sqrt{6}} \left(c_1^{\dagger}- c_3^{\dagger} + c_5^{\dagger} - c_7^{\dagger} + c_9^{\dagger}- c_{11}^{\dagger}\right)\ket{0} \nonumber \\
\ket{E_4}&=&  \frac{1}{2\sqrt{2}} \left(c_1^{\dagger} + c_2^{\dagger} -c_4^{\dagger}-c_5^{\dagger} + c_7^{\dagger}+c_8^{\dagger} - c_{10}^{\dagger} -c_{11}^{\dagger} \right) \ket{0}\nonumber \\
\ket{E_8}&=&  \frac{1}{2\sqrt{2}} \left(c_1^{\dagger} - c_2^{\dagger} +c_4^{\dagger}-c_5^{\dagger} + c_7^{\dagger}-c_8^{\dagger}+ c_{10}^{\dagger} -c_{11}^{\dagger} \right)\ket{0}. \nonumber
\ee
In $\ket{E_6}$, only the odd lattice sites can be occupied  with a constant phase shift $\pi$. This results in a periodic variance by two lattice sites, known  as 2-period. On the other hand, both $E_4$ and $E_8$  displays the periodic occupation in   groups of three consecutive lattice sites, known as 3-period. The periodicity observed in these states is due to the periodic distribution of lattice sites with zero on-site potential, as described by $\Delta_n$ in Eq. \eqref{hs}. This periodicity, along with the presence of quasidisorder, leads to the formation of a mobility edge.  Thus, it  is not surprising that the periodic eigenstate is able to withstand the influence of quasidisorder and remain stable even as the disorder strength approaches infinity, resulting in the mobility edge $E_c^{\infty}$. The stability of these periodic states is protected by the periodicity inherent in the mosaic model. When this periodicity is disrupted, localization occurs, making the poles unstable against dissipation.

Second, there is always an additional pole, such as $z_{10}$ for $\kappa=2$ and $z_9$ for $\kappa=3$ in Table \ref{z}, where the  imaginary part is larger compared  to the others. As demonstrated in Fig. \ref{fig:overlap}, this special mode has a minimal impact on the survival probability of energy level $E_j$ of the initial state. In this case, this mode will serve to characterize the scenario of excitation embedded in the environment, thus not affecting the dissipation of excitation in the system. Similarly, the pole with the largest real part also has a finite imaginary part. This mode can be  considered to be roughly the renormalized highest excited state in the mosaic model, which decays much faster than the other energy levels.

Finally, in Fig.\ref{fig:vN} we observe that the imaginary parts of poles, excluding the additional poles,  exhibit a regular variance with respect to the  mobility edge $E_c$.   For instance,  focusing on the region $\left|\text{Re} z_i \right|\leq 1 $  in the case of $\kappa=2$, the poles in this region generally  have smaller imaginary parts globally compared to  the other poles, as shown in Figs. \ref{fig:vN} (a1). However, we also note   that the poles near $\text{Re}z_i=\pm 2$ can show the much smaller imaginary parts for a special $N$.  Upon further calculation, it is found that their real parts are very closed to the eigenenergy of The mosaic model  in this scenario. Thus, the phenomenon may be attributed to occasional resonance between the dynamical mode and the energy level of the mosaic model, and thus is nonuniversal.   A similar trend can be observed in the case of $\kappa=3$, as shown in  Figs. \ref{fig:vN} (b1).


\begin{figure}
\center
\includegraphics[width=7cm]{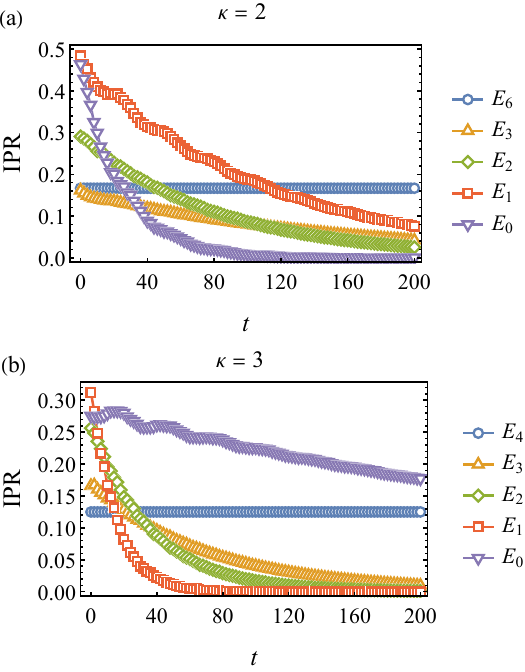}
\caption{(Color online)The time evolution of IPR for four  initial states chosen to be the energy levels in the the mosaic model  in the cases of $\kappa=2$ and $\kappa=3$. The parameters are chosen to be the same as those in Table \ref{z}.   }
\label{fig:t}
\end{figure}

\section{quantum Mpemba effect of localization }

The significant correlation between the dynamical mode and  the localization in the mosaic model can give rise to  unique dynamics of excitation. To measure the localization of the wave function in the system, the inverse participation ratio (IPR)  is introduced, defined as 
\be\label{ipr}
I_{\alpha}= \sum_n \left|\alpha_n\right|^4,
\ee
where $\alpha_n$ denotes the distribution amplitude of the  excitation in the $n$-th site in the system. IPR reaches its minimum  $1/N$ when $\left|\alpha_n\right|^2 = 1/N$ for all $n$, demonstrating a completely extended state. Conversely, the IPR reaches its maximum value of 1 when the excitation is localized on a specific site, indicating complete localization. The time evolution of the IPR  is plotted in Fig. \ref{fig:t} for initial states that are chosen to be the energy levels in  the mosaic model. From the information in Table \ref{z} and Fig. \ref{fig:overlap}, it is evident  that certain  energy levels exhibit significant overlap with special poles, influencing the dissipative dynamics.   As shown in Fig. \ref{fig:t}, IPR shows varying rates of dissipation for different initial states, determined by the imaginary parts of the corresponding poles.  As a result, the  evolution paths  intersect at certain times, indicating the occurrence of  QME \cite{carollo,kochsiek,zhang24}.

It is important to note that  the system can be steady in two different scenarios due to the existence of steady poles. In one scenario,   when excitation is absorbed by the environment, the system is in equilibrium and  IPR tends to be zero, as shown in Figs. \ref{fig:t} (a) and (b) for $E_0, E_1, E_2, E_3$. In the other scenario,   when the system reaches a steady state,  the excitation can remain in the system with a finite probability, and thus IPR can be a finite constant. As shown  in Fig. \ref{fig:t} (a) for $E_6$ and (b) for $E_4$, IPR remains constant as the energy levels  coincide  with the steady pole  $z_6$ for $\kappa=2$ and $z_8$ for $\kappa=3$ respectively. It is worth noting that the QME can be defined only if the initial state with distinct localization decays to the same equilibrium. The presence of steady poles indicates enhanced dissipation due to localization.

Finally, we emphasized that the observed Mpemba effect of localization  is purely quantum. The reason can be illustrated by two facets.  First, it is well known that the localization of the wavefunction in  disordered quantum systems  can be attributed to  the destructive interference in the quantum amplitudes associated with tunneling paths. Thus, the nonequilibrium in the  mosaic model is completely quantum.  Second, since the environment is at zero temperature,  the decay of the excitation  derived from Eq. \eqref{evolution} or \eqref{laplace} is a result of  the coherent exchanging of energy between the system and its environment. Especially, the steady mode can emerge from the inherent periodicity in The mosaic model, which is  protected by the quasidisorder. Deviation from the steady mode can lead to the localization of wave function, enhancing the decay of the excitation.  In short, the Mpemba effect of localization stems from the interplay of  disorder and  dissipation  and thus is purely quantum.

\section{experimental proposition}
 
The experimental realization of  quasiperiodic the mosaic model was proposed recently in Ref. \cite{gao24}.  In this work, the  mosaic lattice can be realized by evanescently coupled waveguides, fabricated on integrated Si$_3$N$_4$ photonics platform. The transport properties    can be studied by measuring the real-space distribution of light intensity  after certain distance of light propagation.  The influence of environment can manifest  as optical transmission losses, which result from scattering, absorption or dispersion during light propagation. In theory, the losses can be quantified by a propagation coefficient $\alpha$, which  represents the exponential decay  of the optical power over distance. This process is physically similar to the spontaneous emission of atomic excitation, which can be captured by a virtual environment with the Lorentz spectral function. By determining the value $\alpha$, one can identify the influence of the energy mode $z$. As for the steady mode,   $\alpha$ may vanish or  be negligible, while a finite $\alpha$ can occur for the other modes. 

In light of the experiment, special focus should be placed on preparing steady modes like $\ket{E_6}$, $\ket{E_4}$, and $\ket{E_8}$ in Sec.III. It is clear that achieving strict periodicity and consistent phase shifts in the steady mode necessitates meticulous preparation of the waveguide structure. Additionally, it is important to note that the quasiperiodic the mosaic model is subject to the periodic boundary condition $c^{(\dagger)}_{N+n}=c^{(\dagger)}_n$, which may pose inconveniences in experimental settings.

\section{Conclusion and discussion}

In conclusion, the single-excitation dynamics in the mosaic model coupled to a bosonic environment was studied analytically in this paper by determining the dynamical modes. We found that the dynamical modes are closely tied  to the energy levels in the mosaic model. Furthermore, due to the existence of robust asymptotic mobility edge, the dynamics of the excitation displays a significant correlation to the localization in the mosaic model.  Especially,   we observe  that  the asymptotic mobility edge $E^{\infty}_c$  corresponds to a steady dynamical mode, which shows vanishing imaginary part, meaning it is  dissipationless. The energy level $E^{\infty}_c$ also exhibits a global periodicity in real configuration,  determined  by the spatial period in the mosaic model. Apart from $E^{\infty}_c$, the dynamical mode shows finite imaginary part, reflecting  the dissipative dynamics. This suggests a relationship between the dissipative dynamics of excitation and the localization properties of the mosaic model. By adopting the IPR as a measure of localization,  the paths of evolution for the IPR  can intersect at different times for various initial states, highlighting the localization-related QME.

Although this study  centered on the single-excitation case, the dynamics for multiple excitations does not exhibit essential differences as a result of the absence of particle interaction. This can be confirmed by observing that the state of multiple excitations can always be expressed as a product form of single excitation states when there is no particle interaction. For convenience, the single excitation is rewritten as 
\be 
\ket{1_n}= \left(\sum_{j=1}^N \alpha_{n, j}  c^{\dagger}_n+ \sum_{k=1}^{K} \beta_{n, k}  b^{\dagger}_k \right)\ket{0} := \hat{e}_n \ket{0}, 
\ee
satisfying $H \ket{1_n}=E^{(1)}_n\ket{1_n}$. With this formula, the total Hamiltonian can be expressed as 
\be 
H=\sum_n E^{(1)}_n  \hat{e}^{\dagger}_n  \hat{e}_n.
\ee
As an example, the  energy state for  double excitations can be written explicitly as 
\be 
\ket{2_n}=\hat{e}_i \hat{e}_j \ket{0}. 
\ee
It is not difficult to get 
\be 
H\ket{2_n} = \left(E^{(1)}_i+E^{(1)}_j \right)\ket{2_n}.
\ee
This observation suggests that the effective mode for double excitations could be a combination of two single excitation modes, leading to an increased decay rate as a sum of the single excitation rates. A state of multiple excitations can be formed in a similar manner.

It should be pointed out that mobility edge can  emerge through the  introduction of short-term \cite{bs10} or long-term hopping \cite{deng19}, or the breaking the duality symmetry \cite{shx88,gps15,lls17,yao19} in the Aubry-Andr\'{e}-Haper model \cite{aa,haper}.  Different from  the The mosaic model, the mobility edges in these cases generally show dependence on the quasidisordered on-site potential, and there is  no  asymptotic mobility edge. Since the occurrence of dissipation is independent of the potential, it is impossible to establish the relationship  between the localization and the decay. Our investigation  of the single excitation dynamics  in the generalized  Aubry-Andr\'{e}-Haper model introduced in Ref. \cite{gps15}  confirms this hypothesis, as shown in \ref{appendix:gaah}.

Finally,  it is crucial for the QME of  the localization that the extended  energy level $E^{\infty}_c$ can withstand dissipation and exhibit the periodicity related to the value of $\kappa$ in the the mosaic model. In physics, the robustness of the energy level  $E^{\infty}_c$ is a result of the homogeneous coupling of the  lattice to the environment, as shown in Eq.\eqref{hint}, keeping the global periodicity in the mosaic model invariant. This suggests that  the asymptotic mobility edge may  persist beyond the single-excitation subspace or in the presence of  particle interaction. However, exploring multiple-excitation dissipative dynamics in interacting many-body systems poses a theoretical challenge that remains to be addressed by  future research.

\section*{ACKNOWLEDGEMENTS}
H.T.C. acknowledges the support of the Natural Science Foundation of Shandong Province under Grant No. ZR2021MA036. M.Q. acknowledges the support of NSFC under Grant No. 11805092 and the  Natural Science Foundation of Shandong Province under Grant No. ZR2018PA012.

\renewcommand\thefigure{A\arabic{figure}}
\renewcommand\theequation{A\arabic{equation}}
\renewcommand\thesection{Appendix \Roman{section}}
\renewcommand\thetable{A\Roman{table}}
\setcounter{equation}{0}
\setcounter{figure}{0}
\setcounter{section}{0}
\setcounter{table}{0}

\section{The localization  of energy level in the the mosaic Model}\label{appendix:E}

\begin{figure}[t]
\center
\includegraphics[width=8.5cm]{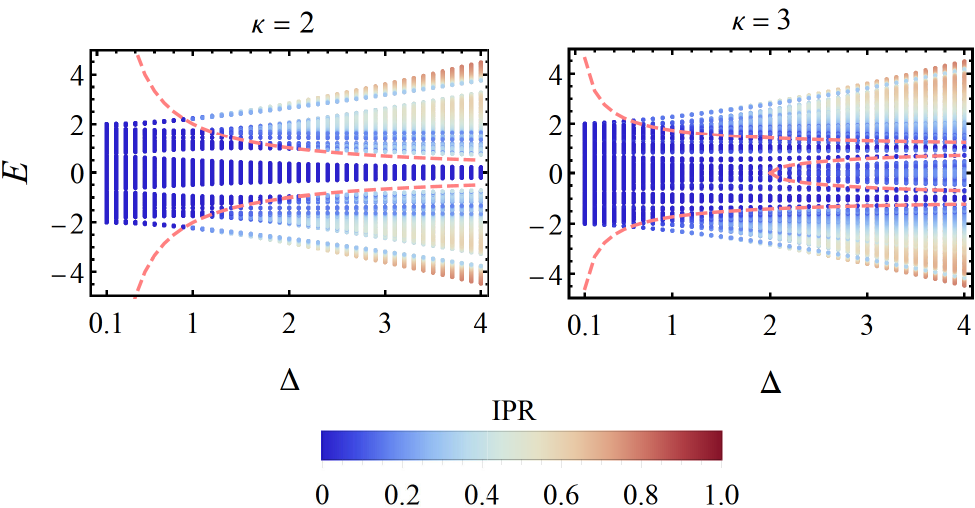}
\caption{(Color online) The plot of IPR for all energy levels in the the mosaic model in the cases of  $\kappa=2$ and $\kappa=3$ respectively. The dashed-pink line characterizes the mobility edge, determined by Eq. \eqref{me}. For the plots, $N=610, \Delta=2$ and $\phi=0$ are chosen.   }
\label{fig:E}
\end{figure}

In this appendix, a brief discussion of the localization in the the mosaic model is presented. It is known that due to the existence of mobility edge, the energy levels in the the mosaic model can exhibit unique localization characteristics. We adopt  the inverse participation ratio defined by Eq. \eqref{ipr} to quantify the localization in the system. As shown in Fig. \ref{fig:E}, the energy level of the system can exhibit either localized or extended behavior depending on its position relative to the mobility edge. The coexistence of localized and extended state is a typical feature for the the mosaic model, leading to the complex dynamics in the system.

The presence of $E_c^{\infty}$ in the the mosaic model is a result of the periodicity, decided by the onsite potential $\Delta_n$ in Eq. \eqref{hs}. It has been observed that $E_c^{\infty}$ is  one of the eigenvalues of $H_s$ only when $N/\kappa$ is even. In physics, the The mosaic model depicts a lattice system that showcases  both disorder  and invariant by discrete translation. It is obvious from the definition of $\Delta_n$ that the disorder presents  only at the site $n=j \kappa$, while the discrete translation invariance occurs for the other sites. The interplay between disorder and discrete translation invariance gives rise to the emergence of the mobility edge.

\begin{table}[tb]
  \centering
  \begin{tabular}{c|c|c}
    \hline \hline
   $z_i$  & $a=0.5$ & $a=-0.5$ \\
     \hline
    $z_1$ & 4.46302 - $\mathbbm{i}$ 0.04866 &2.79814 - $\mathbbm{i}$ 0.21174\\
    $z_2$ &  3.22087 - $\mathbbm{i}$ 0.06612 & 1.72400 - $\mathbbm{i}$  9.101 $\times 10^{-5}$\\
    $z_3$ & 2.27772 - $\mathbbm{i}$ 0.02594 &  1.70546 - $\mathbbm{i}$ 1.913  $\times 10^{-4}$\\
    $z_4$ & 0.41668 - $\mathbbm{i}$ 0.05686  & -0.11475 - $\mathbbm{i}$ 0.006911 \\
    $z_5$ & 0.28352 - $\mathbbm{i}$ 0.05176  & -0.27667 - $\mathbbm{i}$ 0.03026 \\
    $z_6$ &  -0.57714- $\mathbbm{i}$ 0.67695 & -0.68474 - $\mathbbm{i}$ 0.72597 \\
    $z_7$ & -1.70351 - $\mathbbm{i} $ 6.235 $\times 10^{-5} $& -2.45659 - $\mathbbm{i} $ 0.00369 \\
    $z_8$ & -1.85742 - $\mathbbm{i} $ 0.073562 &  -3.47499  - $\mathbbm{i}$ 0.01309\\
    $z_9$ & -2.0735 - $\mathbbm{i} $ 8.954 $\times 10^{-5} $ &  -4.00698  - $\mathbbm{i}$ 0.00802\\
    \hline  \hline
  \end{tabular}
  \caption{The effective mode $z$  for GAAH defined by Eq. \eqref{gaah}. For this evaluation, $N=8, \Delta=2, \phi=\pi, \eta=0.1$, and $\omega_c=1$ are chosen. }\label{gaahz}
\end{table}

\begin{figure}
\center
\includegraphics[width=6cm]{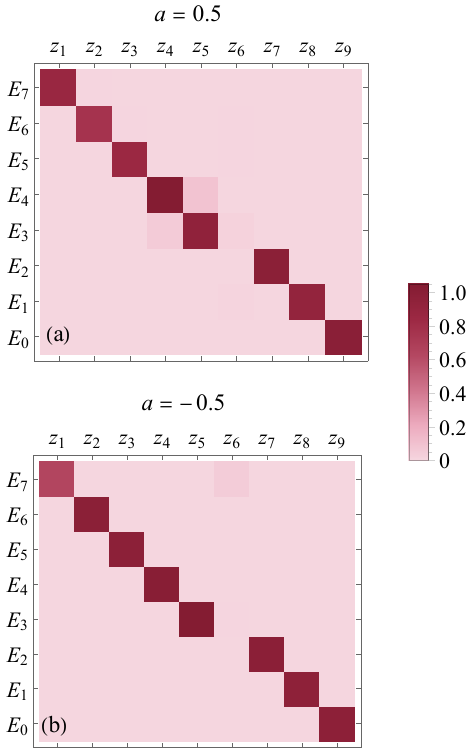}
\caption{(Color online) The square modulus for the overlap between the energy level $E_j (j=0, 1, 2, \cdots, 7 $ in descending order) in the GAAH model  and the coefficient set $\left\{c_{n, i} \right\}(n=1, 2, \cdots, 8)$ related to the initial state $\ket{E_j}$, defined by Eq.(15). For the plots, the parameters are chosen the same as those in Table \ref{z}. For $a=0.5$, $E_j>0$ when $j=3, 4, 5, 6, 7$, while $E_j<0$ otherwise. In contrast, for $a=-0.5$, $E_j>0$ when $j=5, 6,7$, while $E_j<0$ otherwise. }
\label{fig:gaah}
\end{figure}

\section{Effective modes in the generalized Aubry-Andr\'{e}-Harper   model}\label{appendix:gaah}

To demonstrate the special role of $E_{\infty}$, we evaluate the effective modes in the generalized Aubry-Andr\'{e}-Harper (GAAH) model \cite{gps15}, in which the Hamiltonian is 
\be\label{gaah}
H_s=\sum_{n=1}^N c^{\dagger}_{n} c_{n+1} +c^{\dagger}_{n+1} c_{n} +
\frac{\Delta \cos(2\pi \beta n +\phi)}{1 - a \cos(2\pi \beta n +\phi)}c^{\dagger}_n c_{n},
\ee 
where $N$ denotes the number of lattice sites and $c_n (c^{\dagger}_n)$ is the annihilation (creation) operator of the excitation at the $n$-th lattice site. The hopping strength is assumed to be unity. The onsite potential is a smooth function of parameter $a$ in the open interval $a\in(-1, 1)$. When $a=0$, Eq. \eqref{gaah} reduces to the standard  Aubry-Andr\'{e}-Harper model. However, the  GAAH model exhibits an exact mobility edge when $a\neq 0$,  determined by  
\be\label{Ec}
a E_c = \text{sgn}\left(\lambda\right)\left(2\left|\lambda\right| - \left|\Delta\right|\right).
\ee 
Depending on whether $a>0$ or $a<0$, the energy levels display much localization with the increasing or decreasing of $E_j$.  By choosing $\Delta=2$, one gets $E_c=0$.

As an illustrations, we investigate the effective modes for $a=0.5$  and $a=-0.5$ respectively by using the method in Section II.  Furthermore, we also check  the overlap between the coefficient set $\left\{c_{n, i} \right\}(n=1, 2, \cdots, N)$ for  mode $z_i$ and the energy level $E_j (j=0, 1, 2, \cdots, N-1,$ labeled in descending order) in the GAAH model in Fig. \ref{fig:gaah}, which clearly illustrate the relationship between the effective modes and the energy levels  in the GAAH model. Evidently,  for both $a=0.5$ and $-0.5$,  there is an additional mode $z_6$ with a very large imaginary part. Because $z_6$ show negligible overlap with the energy level in the system, it will serve to characterize the scenario of excitation embedding in the environment, thus not affecting the dissipation of the excitation initially in the system.

As for the other modes, there is no evidence for a relationship between the decay rate and localization.  For example,  $z_1$ show a significantly larger decay rate  than those for $z_2$ and $z_3$, as shown for $a=-0.5$ in Table \ref{gaahz}.  Additionally, there is a great overlap with the known extended levels $E_j>0$, shown in Fig. \ref{fig:gaah}(b), indicating that the decay  of excitation  is irrelevant to the localization of system. Similar observation can be made for $z_7, z_8$ and $z_9$ when $a=0.5$, as shown in the second column in Table \ref{gaahz}.

One main difference of the GAAH model from the the mosaic model is the absence of $E_c^{\infty}$. In the GAAH model, $E_c \rightarrow \pm \infty$ as $\Delta \rightarrow \infty$ following Eq. \eqref{Ec},  leading to all energy levels becoming localized. This feature implies that the dissipation induced by coupling to the environment will be irrelevant to the property of localization. Thus, the QME of localization cannot occur in the GAAH model. On the other hand, the The mosaic model includes $E_c^{\infty}$, ensuring the existence of extended states even as $\Delta$ approaches infinity.  Importantly, the energy level $E_c^{\infty}$ is robust against dissipation. This difference allows for a relationship to be established between the decay of excitation and localization by measuring the "distance" from $E_c^{\infty}$. Manifestly, the localization can perfectly serve as the measure.


\begin{thebibliography}{99}

\bibitem{mpemba}
E. B. Mpemba and D. G. Osborne, Cool?, Phys. Educ. {\bf 4}, 172 (1969).

\bibitem{ares23a}
F. Ares, S. Murciano, and P. Calabrese, Entanglement asymmetry as a probe of symmetry breaking, Nat. Commun. {\bf 14}, 2036 (2023).

\bibitem{murciano24}
S. Murciano, F. Ares, I. Klich, and P. Calabrese, Entanglement asymmetry and quantum mpemba effect in the XY spin chain, J. Stat. Mech., 013103 (2024).

\bibitem{ares23b}
F. Ares, S. Murciano, E. Vernier, and P. Calabrese, Lack of symmetry restoration after a quantum quench: An entanglement asymmetry study, SciPost Phys. {\bf 15}, 089 (2023).

\bibitem{chalas24}
K. Chalas, F. Ares, C. Rylands, and P. Calabrese, Multiple crossing during dynamical symmetry restoration and implications for the quantum mpemba effect, J. Stat. Mech. 103101 (2024).

\bibitem{rylands24}
C. Rylands, K. Klobas, F. Ares, P. Calabrese, S. Murciano, and B. Bertini, Microscopic origin of the quantum mpemba effect in integrable systems, Phys. Rev. Lett. {\bf 133}, 010401 (2024).

\bibitem{joshi24}
Lata Kh. Joshi, J. Franke, A. Rath, F. Ares, S. Murciano, F. Kranzl,
R. Blatt, P. Zoller, B. Vermersch, P. Calabrese, Christian F. Roos, and Manoj K. Joshi, Observing the Quantum Mpemba Effect in Quantum Simulations, Phys. Rev. Lett. {\bf 133}, 010402 (2024).

\bibitem{mbl24}
S. Liu, H. -K. Zhang, S. Yin, S.-X. Zhang, and H. Yao, Quantum Mpemba effects in many-body localization systems, arXiv: 2408.07750 (2024).

\bibitem{liu24}
S. Liu, H.-K. Zhang, S. Yin, S.-X. Zhang, Symmetry restoration and quantum Mpemba effect in symmetric random circuits, 	Phys. Rev. Lett. {\bf 133}, 140405 (2024).

\bibitem{chang24}
W.-X. Chang, S. Yin, S.-X. Zhang, and Z.-X. Li, Imgainary-time Mpemba effect in quantum many-body system, arXiv: 2409.06547 (2024).

\bibitem{lu17}
Z. Lu and O. Raz,  Nonequilibrium thermodynamics of the Markovian Mpemba effect and its inverse, Proceedings of the
National Academy of Sciences {\bf 114}, 5083 (2017).

\bibitem{carollo}
F. Carollo, A. Lasanta, and I. Lesanovsky, Exponentially accelerated approach to stationarity in markovian open quantum systems through the mpemba effect, Phys. Rev. Lett. {\bf 127}, 060401 (2021).

\bibitem{kochsiek}
S. Kochsiek, F. Carollo, and I. Lesanovsky, Accelerating the approach of dissipative quantum spin system towards stationarity  through global spin rotations, Phys. Rev. A {\bf 106}, 012207 (2022).

\bibitem{zhang24}
Jie Zhang, Gang Xia, Chun-Wang Wu, Ting Chen, Qian Zhang, Yi Xie, Wen-Bo Su, Wei Wu, Cheng-Wei Qiu, Ping-Xing Chen, Weibin Li, Hui Jing, and Yan-Li Zhou, Observation of quantum strong Mpemba effect, Nat. Commun. {\bf 16}, 301 (2025).

\bibitem{strachan}
D. J. Strachan, A. Purkayastha, and S. R. Clark, Non-Markovian quantum Mpemba effect, arXiv: 2402.05756 (2024).

\bibitem{nava}
A. Nava, adna R. Egger, Mpemba effect in open nonequilibrium quantum systems, Phys. Rev. Lett. {\bf 133}, 136302 (2024).

\bibitem{wang}
X. Wang and J. Wang, Mpemba effect in nonequilibrium opne quantum system, Phys. Rev. Research {\bf 6}, 033330 (2024). 

\bibitem{akc23}
A. K. Chatterjee, S. Takada, and H. Hayakawa, Quantum Mpemba Effect in a Quantum Dot with Reservoirs, Phys. Rev. Lett. {\bf 131}, 080402 (2023).

\bibitem{akc24}
A. K. Chatterjee, S. Takada, and H. Hayakawa, Multiple quantum Mpemba effect: exceptional points and oscillations, Phys. Rev. A {\bf 110}, 022213 (2024).

\bibitem{wsw24}
X. Wang, J. Su and J. Wang, Mpemba meets quantum chaos: Anomalous relaxation and Mpemba crossings in dissipative
Sachdev-Ye-Kitaev models, arXiv: 2410.06669 (2024).

\bibitem{the mosaic}
Y.-P. Wang, X. Xia, L. Zhang, H.-P. Yao. S. Chen, J.-G. You Q. Zhang and X.-J. Liu, One-dimensional quasiperiodic The mosaic Lattice with exact mobility edges,  Phys. Rev. Lett. {\bf 125}, 196604 (2020).

\bibitem{aa} S. Aubry and G. Andr\'{e},  Analyticity Breaking and Anderson Localization in Incommensurate Lattices, Ann. Isr. Phys. Soc. {\bf 3}, 133 (1980);.

\bibitem{haper} P. G. Haper, Single Band Motion of Conduction Electrons in a Uniform Magnetic Field, Proc. Phys. Soc. London Sect. A {\bf 68}, 874 (1955)

\bibitem{bellomo}
B. Bellomo, R. Lo Franco, and G. Compagno, Non-Markovian Effects on the Dynamics of Entanglement, Phys. Rev. Lett. {\bf 99}, 160502 (2007).

\bibitem{bs1}
E. Yablonovitch, Inhibited Spontaneous Emission in Solid-State Physics and Electrincs, Phys. Rev. Lett. {\bf 58}, 2059-2062 (1987).

\bibitem{bs2}
S. John and J. Wang, Quantum Electrodynamics near a Photonic Band Gap: Photon Bound States and Dressed Atoms, Phys. Rev. Lett. {\bf 64}, 2418-2421 (1990).

\bibitem{gao24}
J. Gao, I.M. Khaymovich, X.-W. Wang, Z.-S. Xu, A. Iovan, G. Krishna, J. Jieesi, A. Cataldo, A. V. Balatsky, V. Zwilller, and Ali W. Elsharri, Probing multi-mobility edges in quasiperiodic the mosaic lattices, Science Bulletin, https://doi.org/10.1016/j.scib.2024.09.030.

\bibitem{bs10}
J. Biddle and S. Das Sarma, Predicted Mobility Edges in  One-dimensional Incommensurate Optical latttices: An Exactly Solvable Model of Anderson Localization, Phys. Rev. Lett. {\bf 104}, 070601 (2010).

\bibitem{deng19}
X. Deng, S. Ray, S. Sinha, G. V. Shlyapnikov, and L. Santos, One-Dimensional Quasicrystals with Power-Law Hopping, Phys. Rev. Lett. {\bf 123}, 025301 (2019).

\bibitem{shx88}
S. Das Sarma, Song He, and X. C. Xie, Mobility Edge in a Model One-Dimensional Potential, Phys. Rev. Lett. {\bf 61}, 2144 (1988).

\bibitem{gps15}
S. Ganeshan, J. H. Pixley,  and S. Das Sarma,  Nearest Neighbor Tight Binding Models with an Exact Mobility Edge in One Dimension, Phys. Rev. Lett. {\bf 114}, 146601 (2015).

\bibitem{lls17}
X. Li, X.-P Li, and S. Das Sarma, Mobility edges in one-dimensional bichromatic incommensurate potentials, Phys. Rev. B {\bf 96}, 085119 (2017).

\bibitem{yao19}
H. Yao, A. Khoudli, L. Bresque, and L. Sanchez-Palencia, Critical Behavior and Fractality in Shallow One-Dimensional Quasiperiodic Potentials, Phys. Rev. Lett. {\bf 123}, 070405 (2019).



\end{thebibliography}
\end{document}